\documentstyle[a4,11pt] {article}
\begin{document}

\setlength{\baselineskip}{20pt}

\noindent
{\Large\bf Linear magnetosonic waves in solar wind flow tubes}

\vspace{4mm}

\setlength{\baselineskip}{15pt}

\noindent
{\bf Suresh Chandra, S.V. Shinde, P.G. Musrif and Monika Sharma}

\vspace{2mm}

\noindent
School of Physical Sciences, S.R.T.M. University, Nanded 431 606

\vspace{2mm}

\noindent
Email: suresh492000@yahoo.co.in

\vspace{3mm}

\noindent
{\large The paper was accepted in J. Astrophys. Astron. on 16 August 2007 after proper refereeing and 
now has been rejected on 31 December 2009 by saying that the acceptance was 
provisional. There was no word `provisional' when the paper was accepted.}

\vspace{3mm}

\noindent
{\bf Abstract:}
Nakariakov et al. (1996) investigated the linear magnetosonic waves
trapped within solar wind flow tubes, where they accounted for a slab having
boundaries at $x = \pm d$ and extended up to infinity in the $y$ and $z$
directions.  Srivastava and Dwivedi (2006) claimed to extend that work by
considering a two-dimensional slab. We find that the work of Srivastava and
Dwivedi (2006) is not for a two-dimensional slab and has a number of
discrepancies. Further, their results for body waves are not reliable.

\vspace{2mm}


\section{Introduction}

HELIOS spacecraft observations (Thieme et al., 1990) supported Parker's 
assumption (1963) that the solar wind could be fine-structured in the form of
flow tubes. In these flow tubes, the magnetosonic waves may be excited assuming
the Alfv$\acute{\mbox{e}}$n speed to be less inside the tube than that outside.
Nakariakov et al. (1996) (hereinafter referred to as NRM) investigated 
one-dimensional problem by considering a slab having boundaries at $x = \pm d$ 
and extended up to infinity in the $y$ and $z$ directions.
 Srivastava and Dwivedi (2006) (hereinafter referred to as SD) claimed to 
extend 
the work of NRM by accounting for a two-dimensional slab with a symmetric 
expansion ($\delta$) in the edges of the slab.
They obtained expressions for surface as well as body waves. 

NRM considered a slab having boundaries at $x = \pm d$ and extended up to 
infinity in the $y$ and $z$ directions.
For one-dimensional case, the linearized equations of ideal MHD are (NRM)

\vspace{-3mm}
\begin{eqnarray}
\frac{\mbox{d}^2 V_{xi}}{\mbox{d} x^2} - m_i^2 \ V_{xi} = 0 \label{eq:sd11}
\end{eqnarray}

\noindent
where $i$ is either $o$ (for inside) or $e$ (for outside) the slab. The
transversal
plasma velocity is $V_{xi} \  \mbox{exp} \ i (\omega t - k z)$ and
\begin{eqnarray}
m_i^2 = \frac{[a_{A i}^2 - (a-M_i)^2] [a_{S i}^2 - (a-M_i)^2]}{a_{fi}^2
[a_{T i}^2 - (a-M_i)^2]} \ k^2 \label{eq:sd11a}
\end{eqnarray}

\noindent
where
\vspace{-3mm}
\begin{eqnarray}
a = \frac{\omega}{k \ C_{A o}}  \hspace{2cm} a_{Ai}^2 = \frac{C_{A i}^2}
{C_{A o}^2} \hspace{2cm} a_{Si}^2 = \frac{C_{S i}^2}{C_{A o}^2}
  \nonumber
\end{eqnarray}
\begin{eqnarray}
M_i = \frac{U_i}{C_{A o}} \hspace{2cm} a_{Ti}^2 = \frac{C_{A i}^2 \ C_{S i}^2}
{C_{A o}^2 (C_{A i}^2 + C_{S i}^2)} \hspace{2cm} a_{fi}^2 = \frac{C_{A i}^2 +
 C_{S i}^2}{C_{A o}^2} \nonumber
\end{eqnarray}

\noindent
Here, all the variables are normalized with $C_{A o}$. $M_i$ is
 Alfv$\acute{\mbox{e}}$n Mach number and $a$ the phase speed in the units of 
Alfv$\acute{\mbox{e}}$n 
speed $C_{A o}$. So, NRM accounted for one-dimensional problem where the
thickness $2 d$ of the slab along the $x$-direction is not changing.
When the speed of sound is much larger than all other velocities (the case of
incompressible plasma), the dispersion relation is
\begin{eqnarray}
\frac{\rho_e}{\rho_o} \frac{[a_{Ae}^2 - (a-M_e)^2]}{[1 -
(a-M_o)^2]} = \left\{\begin{array}{rl} - \mbox{tanh}(k d) & \hspace{2mm} \mbox{kink surface waves} \\ - \mbox{coth}
(k d) & \hspace{2mm} \mbox{sausage surface waves} \\ \end{array}\right. \label{eq:naka29a}
\end{eqnarray} 
\noindent
and there are only surface waves.

SD accounted for a two-dimensional slab having boundaries
at $x = \pm d$, $y = \pm d$ at the base and at $x = \pm (d + \delta)$, $y = \pm 
(d + \delta)$ at the top. Later on they converted the expression $\pm (d + 
\delta)$ into $\pm d  \pm  \delta$ without any reason. The later expression
carries some other values in addition to the previous ones and those values are
 irrelevant.

\noindent
For the symmetric expansion in the edges of the slab ($\delta$), the 
conservation of magnetic flux gives
\begin{eqnarray}
B_{0z} \ (2d)^2 = B_z \ (2d + 2 \delta)^2 \nonumber
\end{eqnarray}

\noindent
where $B_z$ and $B_{0z}$ are the magnetic field strengths at the top and the 
base of the slab, respectively. Thus, we have
\begin{eqnarray}
\delta = \left(\frac{B_{0z} - B_z}{B_z}\right) \frac{d}{2} \nonumber
\end{eqnarray}

\noindent
This expression differs from equation (2) of SD and is derived for the situation
that $\delta < < d$. Equation (2) of SD is not even dimensionally correct. It 
is further interesting to find a plus-minus sign in equation (2) of SD, as the sign of $\delta$ is decided by the relative values of $B_z$ and $B_{0z}$. Hence, 
$\delta > 0$, when we have $B_{0z} > B_z$ and for $\delta < 0$, we have $B_{0z} < 
B_z$.  We have taken $B_{0z} > B_z$, so that $\delta$ is positive. Though SD 
claimed a two-dimensional treatment of the problem, but they also used the MHD
equation (\ref{eq:sd11}) which is for a one-dimensional case only. Further, 
SD considered the boundary conditions
\begin{eqnarray}
\frac{V_{xo} (x = \pm d)}{\omega - k U_o} =  \frac{V_{xe} (x = \pm d)}{\omega - 
k U_e}\hspace{1.0cm} \label{eq:sd12}\\ \nonumber\\
pT_o(x = \pm d) = pT_e(x = \pm d)\hspace{1.0cm} \label{eq:sd13}\\ \nonumber\\
pT_i = \frac{i C_{A o} \rho_i a^2_{fi} [a^2_{Ti} - (a - M_i)^2]}{k (a - M_i)
[a^2_{Si} - (a - M_i)^2]} \label{eq:sd14}
\end{eqnarray}

\noindent
where $\rho_i$ is the gas density, which have been used by NRM for 
one-dimensional case. SD did not mention any thing about the $y$ coordinate. It
categorically shows that except giving a figure and 
conservation of magnetic flux, SD did not do any thing with the two-dimensional
case. Their treatment appears as one-dimensional case.

Let us now look into the equations of SD. The equations (\ref{eq:sd12}),
(\ref{eq:sd13}) and (\ref{eq:sd14}) give boundary conditions at $x = \pm d$ and
nothing is said even about the top at $x = \pm (d + \delta)$. Let us assume 
similar boundary conditions at the top also.

\begin{eqnarray}
\frac{V_{xo} [x = \pm (d + \delta)]}{\omega - k U_o} =  \frac{V_{xe} 
[x = \pm (d + \delta)]}{\omega -
k U_e} \label{eq:sd15}\\ \nonumber\\
pT_o[x = \pm (d + \delta)] = pT_e[x = \pm (d + \delta)] 
\label{eq:sd16}
\end{eqnarray}

For the solutions outside the slab, equation (7) of SD should be as the 
following.
\begin{eqnarray}
V_{x e}(x) = \left\{\begin{array}{ll} A_1 \ \mbox{exp}[-m_e \{x - (d+\delta)\}]
 & \hspace{9mm} x > \hspace{4mm} (d+\delta) \\ 
 A_2 \ \mbox{exp}[+m_e \{x + (d+\delta)\}] & \hspace{9mm} x < - (d+\delta) \\
 \end{array}\right.  \label{eq:sd18}
 \end{eqnarray}

\noindent
which correspond to the top of the slab. Here, $A_1$ and $A_2$  are constants.
For the solutions inside the slab, equation (8) of SD should be as the
following.
\begin{eqnarray}
V_{x o} (x) = \left\{\begin{array}{ll}
A \  \mbox{sinh}(m_o x) & \hspace{9mm} \mbox{for sausage surface modes} \\ 
A \  \mbox{cosh}(m_o x) & \hspace{9mm} \mbox{for kink surface modes} \\ 
A \  \mbox{sin}(n_o x) & \hspace{9mm} \mbox{for sausage body modes} \\ 
A \ \mbox{cos}(n_o x) & \hspace{9mm} \mbox{for kink body modes} \\
 \end{array}\right.
 \label{eq:sd19}
 \end{eqnarray}

\noindent
where $n_o^2 = - m_o^2$ and $A$ is a constant. These expressions are the same at the base as well as 
at the top of the slab. On applying boundary conditions (\ref{eq:sd15}) and
(\ref{eq:sd16}) along with  (\ref{eq:sd14}), we get
 for surface waves as 
\begin{eqnarray}
\frac{\rho_e m_o}{\rho_o m_e} \frac{[a_{Ae}^2 - (a-M_e)^2]}{[1 - 
(a-M_o)^2]} = \left\{\begin{array}{r} - \mbox{tanh}[m_o (d + \delta)] \\ - \mbox{coth}
[m_o (d + \delta)] \\
\end{array}\right. \label{eq:sd110}
\end{eqnarray}

\noindent
The upper case corresponds to the kink waves whereas the
lower to the sausage waves. For the body waves, we have 
\begin{eqnarray}
\frac{\rho_e n_o}{\rho_o m_e} \frac{[a_{Ae}^2 - (a-M_e)^2]}{[1 -
(a-M_o)^2]} = \left\{\begin{array}{r} - \mbox{tan}[n_o (d + \delta)]\\ 
\mbox{cot}[n_o (d + \delta)] \\ \end{array}\right. \label{eq:sd111}
\end{eqnarray}

\noindent
Here, also the upper case corresponds to the kink waves whereas the lower to the sausage waves.

\section{Dispersion relations}

Equation (\ref{eq:sd11a}) shows that $m_i$ tends to $k$ in two situations: (i)
when $a = M_o = M_e$, (ii) when the speed of sound is much larger than all other
velocities (the case of incompressible plasma). 

(i) For $a = M_o = M_e$, the steady
shear flows are equal inside as well as outside the slab. Under such situation,
equation.  (\ref{eq:sd110}) reduces to
\begin{eqnarray}
\frac{\rho_e C_{Ae}^2}{\rho_o C_{Ao}^2} = \left\{\begin{array}{rl} - \mbox{tanh}
\{k (d + \delta)\} & \hspace{4mm} \mbox{kink surface waves} \\ 
- \mbox{coth}\{k(d + \delta)\}
& \hspace{4mm} \mbox{sausage surface waves}\\ \end{array}\right. \nonumber
\end{eqnarray}

\noindent
giving no dispersion relation which relates $a$ and $k$.

(ii)
When the speed of sound is much larger than all other velocities (the case of
incompressible plasma), the dispersion relation is
\begin{eqnarray}
\frac{\rho_e}{\rho_o} \frac{[a_{Ae}^2 - (a-M_e)^2]}{[1 -
(a-M_o)^2]} = \left\{\begin{array}{rl} - \mbox{tanh}\{k (d + \delta)\} & 
\hspace{2mm} \mbox{kink surface waves} \\ 
- \mbox{coth} \{k (d + \delta)\} & \hspace{2mm} \mbox{sausage surface waves} \\ 
\end{array}\right.  \label{eq:naka29a}
\end{eqnarray}

\noindent
For $\delta = 0$, these expressions are same as those obtained by Nakariakov et
al. (1996) for one-dimensional case. Moreover, there are only surface waves and 
the body waves do not exist.
 
We could not see any way to replace $n_o$ by $k$ in equation (\ref{eq:sd111}).
Aforesaid expressions show that $n_o$ can be replaced by $i k$, leading to 
non-existence of body waves. But SD have replaced $n_o$ by $k$ in their 
equations (11) and (12) and obtained the expressions
\begin{eqnarray}
\frac{\rho_o}{\rho_e} \frac{[a_{Ae}^2 - a^2]}{[1 -
(a-M)^2]} = \left\{\begin{array}{rl} - \mbox{tan}\{k (d + \delta)\} &
\hspace{2mm} \mbox{kink body waves} \\
 \mbox{cot} \{k (d + \delta)\} & \hspace{2mm} \mbox{sausage body waves} \\
\end{array}\right.  \label{eq:naka29b}
\end{eqnarray}

\noindent
where they taken $M_e = 0$ and $M_o = M$. This expression is  
not correct as $n_o$ cannot be replaced by $k$. 

\section{Conclusions}

The above discussion categorically shows that all the equations of SD are 
objectionable and full of discrepancies. In particular, their expressions for
 When the equations used in the 
calculations are not correct, the results obtained from them cannot be 
reliable.

\section{Acknowledgments}

Financial supports from the Department of Science \& Technology, New Delhi and
the Indian Space Research Organization, Bangalore are thankfully acknowledged.
Thanks are due to the learned referee for encouraging comments.

\end{document}